\documentclass[aps,prb,twocolumn,amsfonts,amsmath,amssymb,superscriptaddress]{revtex4-1}
\usepackage{graphicx}
\usepackage{dcolumn}
\begin{document}

% Title of the article
\title{Observation of anisotropic interlayer Raman modes in few-layer ReS$_2$}
\author{Philipp Nagler}
\affiliation{Institut f\"ur Experimentelle und Angewandte Physik,
Universit\"at Regensburg, D-93040 Regensburg, Germany}
\author{Gerd Plechinger}
\affiliation{Institut f\"ur Experimentelle und Angewandte Physik,
Universit\"at Regensburg, D-93040 Regensburg, Germany}
\author{Christian Sch\"uller}
\affiliation{Institut f\"ur Experimentelle und Angewandte Physik,
Universit\"at Regensburg, D-93040 Regensburg, Germany}
\author{Tobias Korn}
\email{tobias.korn@physik.uni-regensburg.de}
\affiliation{Institut f\"ur Experimentelle und Angewandte Physik,
Universit\"at Regensburg, D-93040 Regensburg, Germany}

\begin{abstract}
ReS$_2$ has recently emerged as a new member in the rapidly growing family of two-dimensional materials. Unlike MoS$_2$ or WSe$_2$, the optical and electrical properties of ReS$_2$ are not isotropic due to the reduced symmetry of the crystal. Here, we present layer-dependent Raman measurements of ReS$_2$ samples ranging from monolayers to ten layers in the ultralow frequency regime. We observe  layer breathing  and  shear modes which allow for easy assignment of the  number of layers. Polarization-dependent measurements give further insight into the crystal structure and reveal an energetic shift of the shear mode which stems from the in-plane anisotropy of the shear modulus  in this material.
\end{abstract}
\maketitle   % please do not remove

In the emerging field of two-dimensional (2D) materials science, reports of atomically thin ReS$_2$~\cite{Tongay2014} have lately attracted increasing interest in the research community. ReS$_2$ is a layered semiconductor with a bulk band gap of about 1.5~eV, where one monolayer (ML) consists of one lattice of rhenium atoms which is sandwiched between two lattices of sulfur atoms. Like in MoS$_2$ or WSe$_2$, the weak van der Waals (vdW) interaction between  individual MLs enables the fabrication of atomically thin samples by mechanical exfoliation~\cite{Novoselov2005}. However, in contrast to many other transition metal dichalcogenides (TMDCs), ReS$_2$ and ReSe$_2$ do not crystallize  in a 2H structure but in a distorted 1T structure due to charge decoupling from an additional valence electron in the rhenium atoms~\cite{Tongay2014,Horzum2014} which leads to a reduced crystal symmetry. Recently, it has also been proposed that ReS$_2$ is not isostructural to ReSe$_2$, but crystallizes in a different structure containing only a single ReS$_2$ layer per unit cell~\cite{Feng2015}.  The  in-plane anisotropy of ReS$_2$ has already been examined in electrical and optical studies on bulk samples in the past~\cite{Friemelt1993,Bobet2011}. Very recently, these experiments were extended to atomically thin ReS$_2$ and have demonstrated the potential of the material for future ultra-thin optoelectronic devices~\cite{Corbet2015,Lin2015a,Lui2015,Liu2015}. These hopes are also nurtured by alternative production methods for atomically thin ReS$_2$ such as chemical exfoliation~\cite{Fujita2014} and CVD growth~\cite{Keyshar2015}.

Raman spectroscopy has been established as a powerful tool to study elementary excitations in layered 2D materials by optical means~\cite{Ferrari2013,Zhang2015}. First Raman studies on atomically thin ReS$_2$ in the high-frequency range (\textgreater 100\,$\textrm{cm}^{-1}$) have already provided insight into the intralayer modes~\cite{Feng2015} and their anisotropic behaviour~\cite{Chenet2015}. However, in addition to the intralayer modes, the weak vdW coupling in layered crystal structures usually leads to the emergence of interlayer phonon modes which are typically very low in energy (\textless 50\,$\textrm{cm}^{-1}$). Thereby, each layer oscillates rigidly as a whole unit which leads to two distinct features:  layer shear modes (LSM), where the layers oscillate rigidly against each other in the plane of the crystal, and  layer breathing modes (LBM), where the oscillation amplitude is perpendicular to the layer plane. Pioneered for graphene~\cite{Tan2012}, the study of interlayer phonon modes in atomically thin layered materials was soon extended to TMDCs such as MoS$_2$~\cite{Plechinger2012,Zhang2013}, WSe$_2$~\cite{Zhao2013b}, WS$_2$~\cite{Chen2015} and MoTe$_2$~\cite{Froehlicher2015}. The observed interlayer phonon modes provide valuable information about the  forces acting in the crystal and their layer-dependent behavior renders them a favorable tool for determining the thickness of the material. However, an investigation of the low-energetic interlayer modes of ReS$_2$ is still lacking. It was also argued that for the  proposed bulk crystal structure containing only a single ReS$_2$ layer per unit cell~\cite{Feng2015}, the interlayer phonon modes should not be present at all in ReS$_2$.

Here, we aim to clarify this issue by presenting layer-dependent and polarization-resolved Raman studies of ReS$_2$ in the ultralow frequency regime (\textless 50\,$\textrm{cm}^{-1}$). We clearly observe the LSM and LBM for bilayers and  thicker samples. The layer-dependent peak positions enable a reliable determination of the thickness of ReS$_2$ by Raman spectroscopy. Angle-dependent measurements demonstrate the anisotropic Raman intensity of the interlayer modes and reveal a lifting of the degeneracy of the LSM, which is expected for crystals with anisotropic shear modulus.

\begin{figure}[htb]%
	\includegraphics*[width= \linewidth]{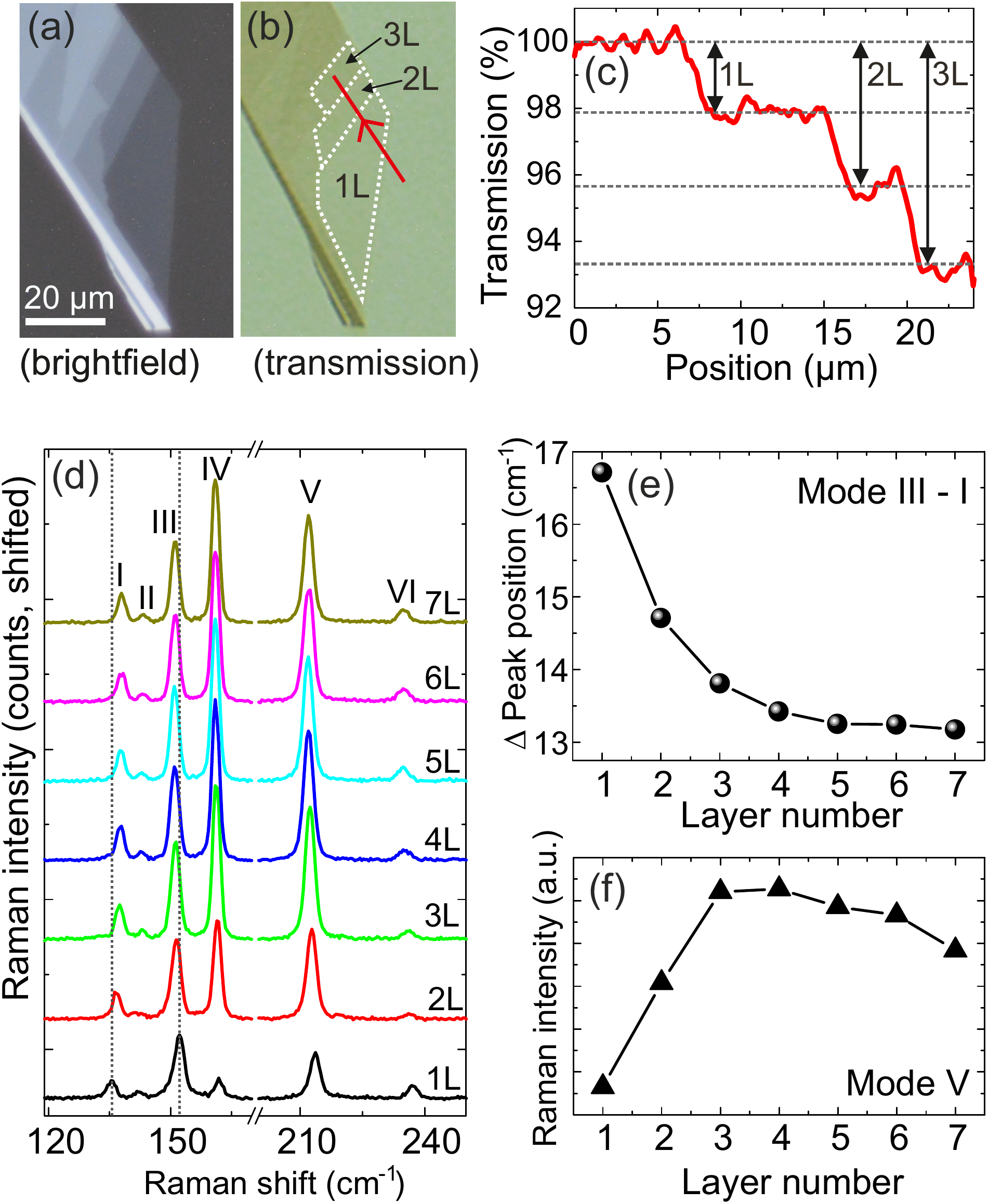}
	\caption{(a) Brightfield and (b) transmission optical microscope images of a ReS$_2$ flake on a PDMS substrate.  The dotted lines in (b) highlight areas of different thickness. The red arrow indicates the position of the linescan shown in (c). (c) Normalized transmission trace extracted from the red channel of (b). (d) Layer-dependent high-energy Raman spectra of ReS$_2$. The dotted lines indicate the peak positions of mode I and III, respectively, in the ML case. (e) Peak frequency difference of mode III and mode I in dependence of the layer number. (f) Raman intensity of mode V in dependence of the layer number.}
	\label{fig:Plot1}
\end{figure}

Our sample preparation starts with mechanical exfoliation of ReS$_2$ bulk crystals (source: 2D Semiconductors Inc.) onto a polydimethylsiloxane (PDMS) stamp. Relevant flakes are identified using an optical microscope and subsequently transferred on a p-doped Si-Chip with a 285\,nm SiO$_2$ capping layer by applying a recently developed all-dry transfer technique~\cite{Castellanos-Gomez2014}. None of the samples showed indications of degradation when stored in ambient conditions for several weeks.
Raman measurements of the samples are carried out at room temperature in backscattering geometry with a linearly polarized 532\,nm cw laser which is coupled into a 100x microscope objective (spot size 1 $\mu$m), further experimental details are published elsewhere~\cite{Plechinger_2DMat}.
All Raman measurements are performed in cross-polarized geometry to suppress the spectrally broad background of inelastically scattered light which stems from free carriers in the heavily p-doped Si substrate~\cite{pSi_Cardona}. To access the ultralow frequency regime we use three reflective volume Bragg grating filters which are placed in front of the spectrometer. Careful adjustment of the setup allows the simultaneous observation of Stokes and Anti-Stokes Raman signals and detection of Raman peaks down to 5\,$\textrm{cm}^{-1}$. For angle-resolved Raman measurements we rotate the sample on a rotary stage while keeping the linear polarization of the laser and the orientation of the polarizer in front of the spectrometer fixed.

First, we  establish a basic understanding of the layer thickness of our samples. Fig.~\ref{fig:Plot1}(a) shows a brightfield image of a ReS$_2$ flake which has been mechanically exfoliated on a PDMS stamp. Terraces of different thickness of the crystal can be  identified but no clear statement about the absolute layer thickness can be made from these brightfield images. However, it has been shown that the absorption of atomically thin materials is fundamentally related to the fine structure constant with one layer absorbing around 2\% of the incident light~\cite{Nair2008,Fang2013}.
Therefore, we examine the same flake in a transmission microscope (Fig.~\ref{fig:Plot1}(b)) and extract the absolute amount of absorbed light from the red channel of the image (Fig.~\ref{fig:Plot1}(c)). The well-defined steps in the transmission of the flake with values of around 2\% for each individual terrace demonstrate that the sample is indeed atomically thin with areas down to the ML limit.

Furthermore, recent Raman measurements have shown that, in analogy to MoS$_2$~\cite{Lee2010}, the energetic difference of certain high-energy modes of ReS$_2$ can be used to assign the thickness of ReS$_2$ flakes from one layer up to four layers~\cite{Chenet2015}. Fig.~\ref{fig:Plot1}(d) displays the layer-dependent high-energy Raman modes of ReS$_2$ up to 7 layers. In line with recent literature~\cite{Feng2015,Chenet2015} we label the individual peaks with Roman numerals from I to VI. For the respective peak positions in the case of a ML we measure: 135.5\,$\textrm{cm}^{-1}$, 142.2\,$\textrm{cm}^{-1}$, 152.2\,$\textrm{cm}^{-1}$, 161.9\,$\textrm{cm}^{-1}$, 213.6\,$\textrm{cm}^{-1}$ and 237.1\,$\textrm{cm}^{-1}$, which is in good agreement with references~\cite{Feng2015} and~\cite{Chenet2015}. Mode I and mode II are assigned to A$_g$-like vibrational modes, while  modes III-VI are E$_g$-like vibrational modes~\cite{Feng2015}. The peak position of mode I increases from 135.5\,$\textrm{cm}^{-1}$ in the ML to 137.6\,$\textrm{cm}^{-1}$ for a 4L sample. By contrast, the position of mode III decreases from 152.2\,$\textrm{cm}^{-1}$ in the ML to 151.1\,$\textrm{cm}^{-1}$ for a 4L sample. Hence, the difference of the peak positions of mode III and mode I provides a clear indication for the thickness of ReS$_2$ up to four layers (Fig.~\ref{fig:Plot1}(e)) ranging from 16.7\,$\textrm{cm}^{-1}$ in the ML to 13.4\,$\textrm{cm}^{-1}$ in the 4L. However, from five layers on, the peak difference converges around 13.2\,$\textrm{cm}^{-1}$ and is therefore no longer an unambiguous measure for the layer thickness. We also determine the absolute Raman signal for each mode with respect to the layer thickness. Figure~\ref{fig:Plot1}(f) shows the Raman intensity of mode V from 1L to 7L. One can observe that the absolute Raman signal increases from 1L on and reaches its maximum in the region of 3L to 4L. This layer-dependent intensity behavior was also observed for the closely related ReSe$_2$~\cite{Wolverson2014a} and further supports our assignment. This effect can be understood by taking into account the optical interference effects between the atomically thin material and the SiO$_2$/Si-substrate which lead to a maximum in the enhancement factor for three layers~\cite{Li2012}.

We now turn to  the interlayer Raman modes, which are found in the ultralow frequency regime. Generally, for an N-layered crystal there are N-1 LBMs perpendicular to the sample plane and N-1 LSMs along the crystal plane~\cite{Michel2012}.  In isotropic crystals, such as graphene or MoS$_2$, the LSMs are doubly degenerate. It is expected, however, that this degeneracy is lifted in materials with in-plane anisotropy~\cite{Zhao2015}. The frequencies of the LBM and LSM as a function of N can be well-described  using a monoatomic chain model~\cite{Zhang2013}.  Thereby, the individual MLs are considered as single mass units which are held together by weak interlayer bonds. This model yields a very simple result for the highest- ($\omega_+$) and lowest-energy ($\omega_-$) LBM and LSM for a given number of layers:
\begin{equation}
\omega_\pm(N)=\omega(2)\sqrt{1 \pm \cos\left(\frac{\pi}{N}\right)},
\label{MCM}
\end{equation}
where $\omega(2)$ are the  frequencies of the LBM and LSM for 2L, respectively.

\begin{figure}[hhh]
	\includegraphics*[width=\linewidth]{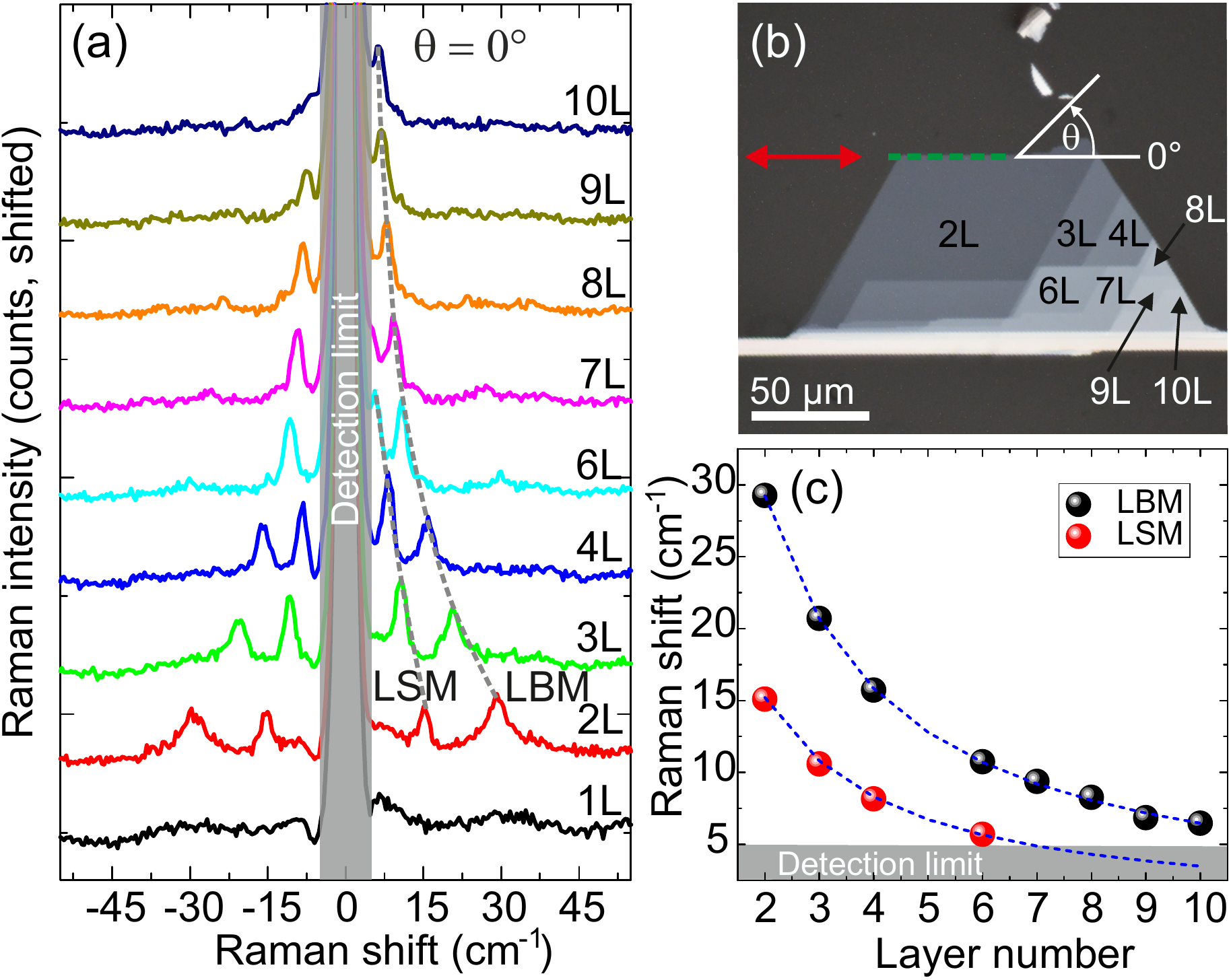}
	\caption{(a) Layer series of Raman spectra of ReS$_2$ in the ultralow frequency regime. The dotted lines are guides to the eye. (b) Brightfield image of the few-layer ReS$_2$-flake (on the PDMS film before transfer) on which the spectra shown in (a) were obtained. The red double arrow shows the orientation of the linearly polarized laser. The green dotted line defines the cleaved edge of the crystal. The white angle shows the direction of rotation and the definition of the angle $\theta$. (c) Layer-dependent peak positions of LBM (black spheres) and LSM (red spheres). The blue dotted lines are fit functions which result from the applied monoatomic chain model.}
	\label{fig:Plot2}
\end{figure}

Figure~\ref{fig:Plot2}(a) shows the layer-dependent Raman spectra of ReS$_2$ ranging from 1L to 10L close to the laser line, which were obtained on the sample shown in Fig.~\ref{fig:Plot2}(b). For these measurements, a cleaved edge of the sample was aligned parallel to the linear polarization of the laser, corresponding to a rotation angle of $\theta=0^\circ$ (see Fig.~\ref{fig:Plot2}(b)).
As expected for any layered material we do not observe  interlayer modes in the ML. In the case of a bilayer we identify two features at 15.1\,$\textrm{cm}^{-1}$ and 29.3\,$\textrm{cm}^{-1}$, which we relate to the LSM and LBM, respectively. Angle-resolved Raman measurements of both modes will later confirm this assignment. The LSM and LBM apparently possess a considerable coupling strength which renders their relative intensity comparable to the features of the high-energy Raman modes (see Fig. 1 of the supporting information). For all layers, we  observe only the energetically lowest ($\omega_-$) modes of the LSM and LBM, albeit N-1 modes are theoretically allowed. This observation might be related to a vanishing electron-phonon coupling strength of the higher-lying modes. In contrast to, e.g., MoS$_2$, where the LBM is suppressed in cross-polarized backscattering geometry due to symmetry reasons~\cite{Zhang2013}, we clearly observe it in ReS$_2$ with a similar amplitude as the LSM.

Figure~\ref{fig:Plot2}(c) shows the peak position of both LSM and LBM with respect to the number of layers. The values were obtained by using Lorentz-shaped fit functions.  Both modes show a softening with increasing N, the LBM from 29.3\,$\textrm{cm}^{-1}$ in 2L to 6.5\,$\textrm{cm}^{-1}$ in 10L and the LSM from 15.1\,$\textrm{cm}^{-1}$ in 2L to 5.7\,$\textrm{cm}^{-1}$ in 6L. For larger N, the LSM shifts below the detection limit. While this behavior is typically observed for the LBM in other layered materials, we point out here that only a few reports exist of the softening branch of the LSM~\cite{Zhang2013,Zhao2015}. By contrast, the most prominent LSM  in graphene~\cite{Tan2012} and MoS$_2$~\cite{Plechinger2012,Zhang2013} is the $\omega_+$ mode that stiffens with increasing N. This stiffening branch of the LSM is not visible in our measurements, which is in line with a recent study on the closely related ReSe$_2$~\cite{Zhao2015}.
In order to fit the extracted peak positions of LBM and LSM we resort to the monoatomic chain model described above, using the layer dependence of the lowest-energy modes $\omega_-$. As can be seen in Fig.~\ref{fig:Plot2}(c) the model quantitatively describes the peak positions of both interlayer modes.
We therefore conclude that the peak positions of the interlayer modes present an unambiguous and destruction-free method to determine the thickness of ReS$_2$ for up to ten layers.

Finally, we present our results of angle-resolved Raman measurements of ReS$_2$ in the ultralow frequency range. As discussed above, we define the rotation angle  $\theta=0^\circ$ for the incident laser polarization being aligned parallel to the cleaved edge of our sample indicated in Fig.~\ref{fig:Plot2}(b). It was suggested recently that well-defined edges of mechanically exfoliated ReS$_2$ often represent the $b$-axis of the crystal~\cite{Chenet2015}. Thereby it was also shown that the respective cleaved edge lies exactly between the intensity lobes of the high-energetic mode V in cross-polarization geometry.
\begin{figure}[hhh]
	\includegraphics*[width=\linewidth]{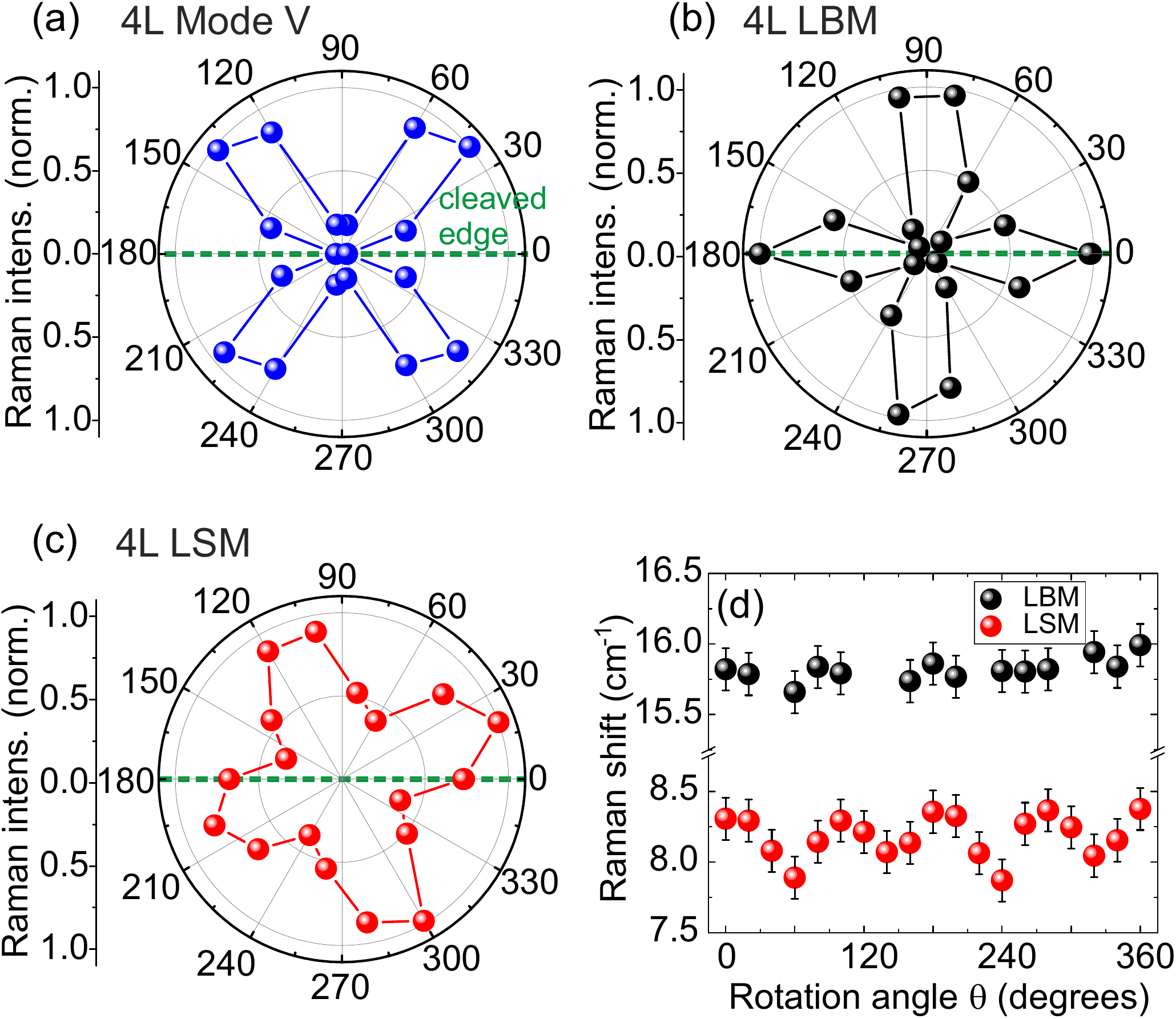}
	\caption{(a) Angle-resolved Raman peak intensity of the mode V in 4L ReS$_2$ located at 212.0\,$\textrm{cm}^{-1}$.  (b) Angle-resolved Raman peak intensity of the LBM in 4L ReS$_2$. (c) Angle-resolved Raman peak intensity of the LSM in 4L ReS$_2$. The maximum of intensity is shifted by 20$^\circ$ with respect to the LBM. The cleaved edge of the sample defined in Fig.~\ref{fig:Plot2}(b) is marked by the green dotted lines in (a)-(c). The  peak intensities shown in (a)-(c) are normalized to the maximum Raman peak intensity of the respective modes. (d) Peak positions of LBM (black spheres) and LSM (red spheres) as a function of the rotation angle. The error bars indicate the spectral resolution of the setup.}
	\label{fig:Plot3}
\end{figure}
Comparing this observation with our results, Fig.~\ref{fig:Plot3}(a) shows the polarization-resolved Raman measurements of mode V in 4L-ReS$_2$. Since the cleaved edge of the crystal lies also exactly between the main lobes of the signal we  assign this direction to the crystal $b$-axis. Figure~\ref{fig:Plot3}(b) and (c) display the angle-resolved Raman measurements of the LBM and LSM in 4L ReS$_2$. Both modes clearly exhibit an anisotropic behavior with four intensity maxima which is a direct consequence of the crystal anisotropy of ReS$_2$. In the case of the LBM the maxima at 0$^\circ$ and 180$^\circ$ lie directly parallel to the cleaved edge of the sample which can serve as an additional fingerprint for the crystal orientation of ReS$_2$. We note that the angles of maximum intensity of the LSM are shifted by 20$^\circ$ with respect to the LBM, which highlights the different nature of these two low-energetic features.

Furthermore, it is also expected that the energetic positions of LBM and LSM show a differing behavior in angle-resolved measurements. Since ReS$_2$ possesses an in-plane anisotropy, the degeneracy of the LSM energy should be lifted as it represents a rigid in-plane vibration. The energy of the LBM however, being an out-of-plane vibration, should naturally not be affected by the in-plane anisotropy. Figure~\ref{fig:Plot3}(d) displays the measured peak position of LSM and LBM for 4L ReS$_2$ in dependence of the rotation angle. The energy of the LSM shows a periodic oscillation with a minimum around 7.8\,$\pm\,0.3\,\textrm{cm}^{-1}$ and a maximum of 8.4\,$\pm\,0.3\,\textrm{cm}^{-1}$ yielding an energetic shift of about 0.6\,$\pm\,0.3\,\textrm{cm}^{-1}$. This angle-dependent shift indicates that the interlayer shear modulus is anisotropic. Using the monoatomic chain model~\cite{Tan2012,Zhang2013},  we can estimate that this anisotropy is on the order of 14~\%. Recently, a theoretical calculation predicted a splitting of the LSM in the structurally closely related ReSe$_2$ in the order of 1\,$\textrm{cm}^{-1}$~\cite{Zhao2015} which is in good agreement with our obtained value for ReS$_2$. In contrast to the LSM, the energy of the LBM does not exhibit any systematic variation with the rotation angle. This observation  corroborates our prior spectral assignment of the two peaks. We also note that none of the high-energy Raman modes show a systematic frequency shift with the rotation angle (see Fig. 2 of the supporting information), in agreement with the study by Chenet et al.~\cite{Chenet2015}, demonstrating the unique sensitivity of the LSM to interlayer coupling anisotropy.

In conclusion, we have presented Raman measurements of ReS$_2$ in the ultralow frequency range. Thereby, we were able to identify and track layer-dependent features of the LBM and the LSM which stem from the interlayer interaction of the crystal. The presence of these modes indicates that ReS$_2$ is isostructural to ReSe$_2$. The layer dependence of their peak positions can be readily reproduced by means of a  monoatomic chain model and provide an easy and non-destructive method for determining the layer number of ReS$_2$ up to 10L. Angle-resolved Raman measurements clearly confirm the anisotropic nature of both interlayer modes. By tracing the peak position of the LSM with respect to the polarization plane we are able to unveil an energetic shift of about 0.6\,$\textrm{cm}^{-1}$ which is directly related to the in-plane shear modulus anisotropy of the crystal. Our Raman measurements on ReS$_2$ therefore should foster the understanding of the fascinating anisotropic properties of this novel 2D material.
\section*{Acknowledgements}
The authors ackowledge financial support by the DFG via KO3612/1-1, GRK1570 and SFB689.

\bibliography{library}
\onecolumngrid
\appendix
\subsection{Supplementary note 1: direct comparison of low- and high-energy Raman modes}
\begin{figure}[h]
\includegraphics[width=0.5 \linewidth]{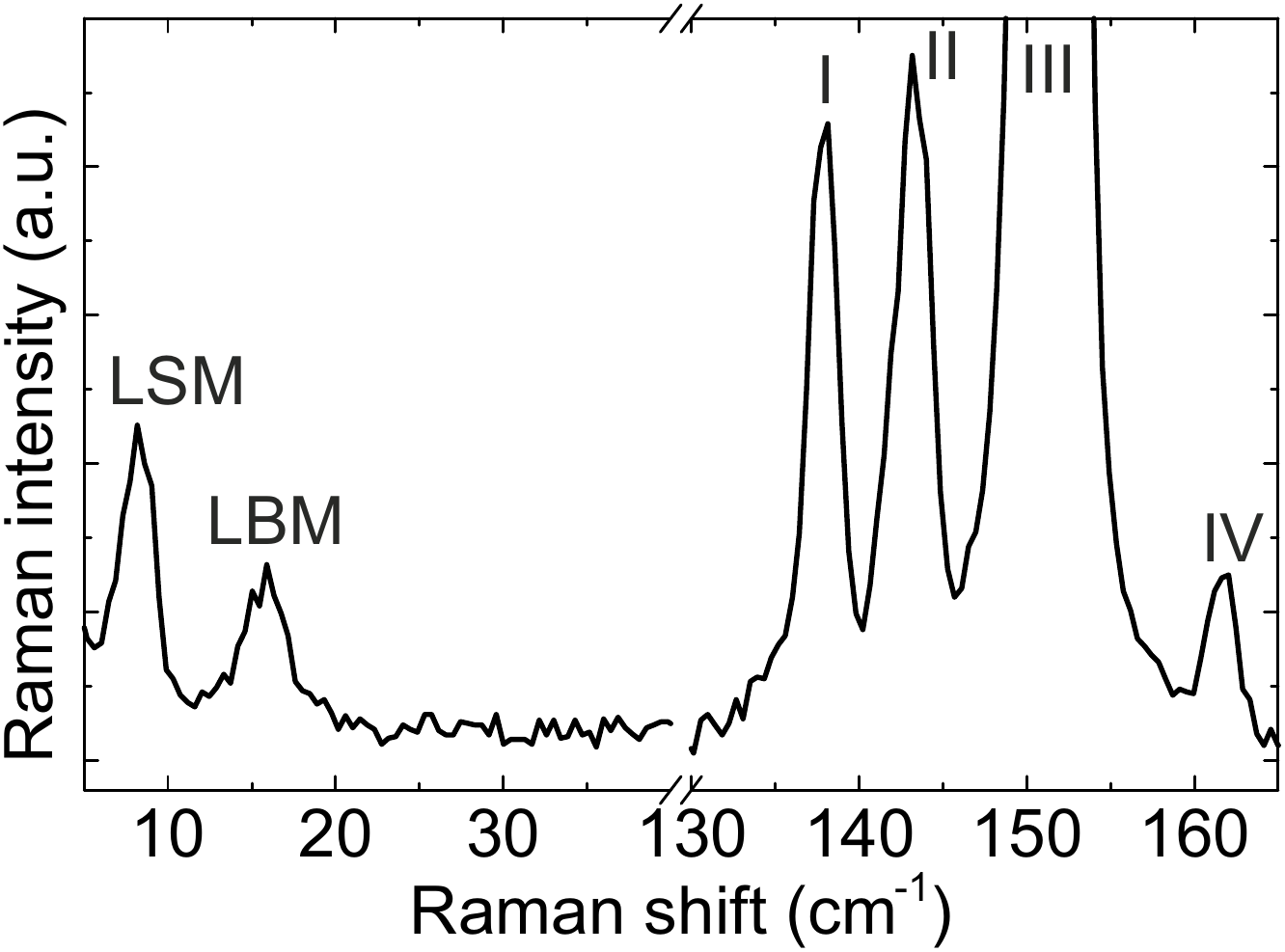}
\caption{Raman spectrum of 4L ReS$_2$ showing the low-energy region relevant for LBM and LSM, as well as the high-energy region containing modes I-IV. The y axis is the same for the two regions.}
\label{Supp1}
\end{figure}

The LSM and LBM have Raman intensities that are on the same order of magnitude as the higher-energy modes. Figure~\ref{Supp1} demonstrates this with a Raman spectrum obtained on 4L ReS$_2$.
\subsection{Supplementary note 2: Mode frequencies of higher-energy modes as a function of angle}
None of the high-energy Raman modes in ReS$_2$ show a systematic frequency shift with the rotation angle, in contrast to the LSM (Fig. 3(d) in the main text). Figure~\ref{Supp1} demonstrates this for the high-energy modes I-V in 4L ReS$_2$. The mode position values were obtained by using Lorentz-shaped fit functions.
\begin{figure}[h]
\includegraphics[width=0.5 \linewidth]{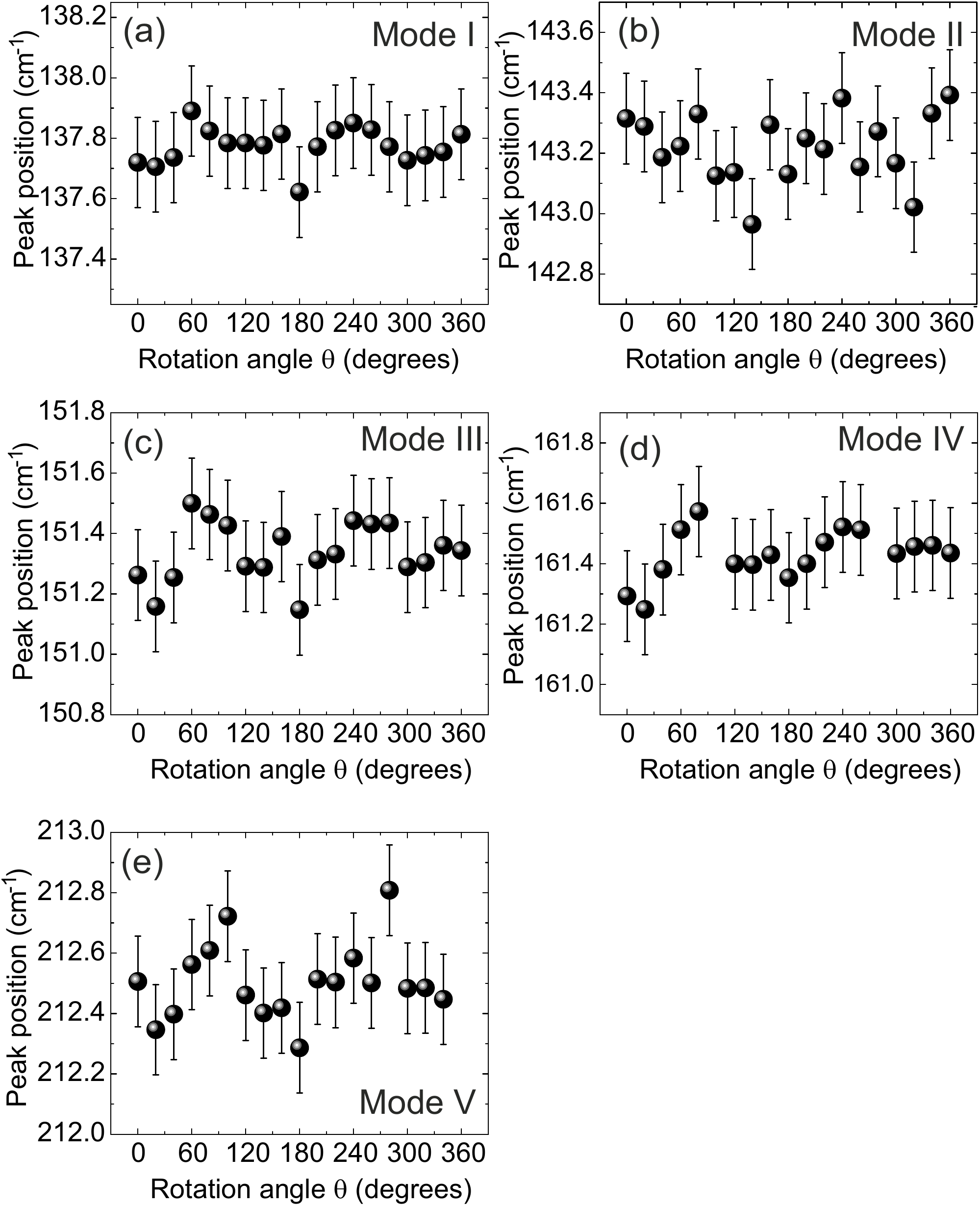}
\caption{(a)-(e) Mode frequencies of the high-energy modes I-V extracted from angle-resolved Raman measurements on  4L ReS$_2$. The y axis covers the same frequency range for all graphs.}
\label{fig:SuppGate}
\end{figure}
\end{document}